# Analyzing Interstellar Infrared Spectrum by Circumcoronene (C54H18) Related Molecules

NORIO OTA

Graduate School of Pure and Applied Sciences, University of Tsukuba,
1-1-1 Tenoudai Tsukuba-city 305-8571, Japan;   n-otajitaku@nifty.com

It is very important to identify carrier molecules of astronomical interstellar infrared spectrum (IR) to understand chemical evolution step of polycyclic aromatic hydrocarbon (PAH) in the universe. In our previous study, it was suggested that coronene ($C_{24}H_{12}$) related PAH could reproduce such IR by the first principles quantum chemical calculation. In this paper, PAH candidates were enhanced to circumcoronene ($C_{54}H_{18}$) related molecules. Well known PAH oriented wavelength of 6.2, 7.7, 8.6, 11.3, and 12.7 micrometer were reproduced well by a void induced mono-cation ($C_{53}H_{18}$) having hydrocarbon two pentagons combined with 17 hexagons. Typical astronomical object are the red rectangle nebula and NGC6946, which is categorized as Type-B sectrum. Neutral circumcoronene show very strong IR peak at 11.1 micrometer, which is noted as Type-A spectrum observed in NGC1316 and NGC4589. Ubiquitously well observed spectrum was noted as Type-C, which could be explained by a suitable combination of Type-B and Type-A. Type-C objects are NGC7023, NGC2023 and so many. It should be noted that a dehydrogenated pure carbon mono-cation molecule ($C_{53}$) show IR peaks at 6.3, 7.8, 8.5 micrometer, but no peak at 11.3 micrometer, which newly defined as Type-E. Such a strange characteristic can contribute to explain IR intensity ratio. Observed intensity ratio between peaks of 6.2, 7.7, 8.6, 11.3, and 12.7 micrometer each other were compared with above calculated IR intensities. Major intensity ratio was well reproduced by Type-B molecules. Variation of observed intensity ratio could be explained by a mixture degree between Type-B and Type-A, and also a mixture degree between Type-B and Type-E.

## 1, INTRODUCTION

It is very important to identify carrier molecules of astronomical interstellar infrared spectrum (IR) to understand chemical evolution step of polycyclic aromatic hydrocarbon (PAH) in the universe. In our previous study, it was suggested that coronene (C24H12) related PAH could reproduce such IR by the first principles quantum chemical calculation. Interstellar infrared spectrum (IR) due to polycyclic aromatic hydrocarbon (PAH) was ubiquitously observed (Boersma et al. 2013, 2014). However, any single PAH molecule or related species showing universal infrared spectrum had not yet been identified. In 2014, it was found for the first time that coronene related molecule $(C_{23}H_{12})^{2+}$ shows very similar infrared spectrum with observed one (Ota 2014b, 2015a, 2017a). This molecule contains two hydrocarbon pentagons combined with five hexagons. This may lead to an identification of specific carrier molecule. Also, simpler molecule $(C_{12}H_8)^{3+}$ had also show good coincidence with observed strong bands, which configuration was hydrocarbon one pentagon combined with two hexagons (Ota 2015b), which may be one candidate to become a basic material to evolve to creation of life itself (Ota 2016).  Based on observation of nebula NGC 2023 (Peeters 2017), we could find out that among 16 observed bands, 14 bands were successfully reproduced well by calculation (Ota 2017b) including IR intensity ratio (Ota 2017c). Recently, it was calculated that dehydrogenated pure carbon molecules also plays an important role on interstellar IR (Ota 2017d).
  In this paper, in order to understand the selection rule of interstellar PAH, molecular candidates were enhanced from coronene to larger size molecule as like circumcoronene (C54H18).

## 2, MODEL MOLECULES AND CALCULATION STEP

In Figure 1(1), primary model molecule is illustrated as photo-ionized cationic circumcoronene $(C_{48}H_{18})^{n+}$ (n=0, +1, +2, and +3). It was supposed that high speed proton from a central star may sputter such molecules and make a carbon void. Once void is created at a position at c or d in (1), there occurs quantum chemical configuration change as illustrated in (2) as void-induced configuration change. This molecule $(C_{53}H_{18})^{n+}$ has hydrocarbon two pentagons combined with 17 hexagons. Mechanism of configuration change was discussed previously by Jahn-Teller deformation in graphene (Ota 2014a). If particle sputter and/or photon irradiation would be more serious, peripheral hydrogen will be removed and will be transformed to pure carbon molecule $(C_{53})^{n+}$ as shown in (3), which configuration is carbon one pentagon combined with 18 hexagons.

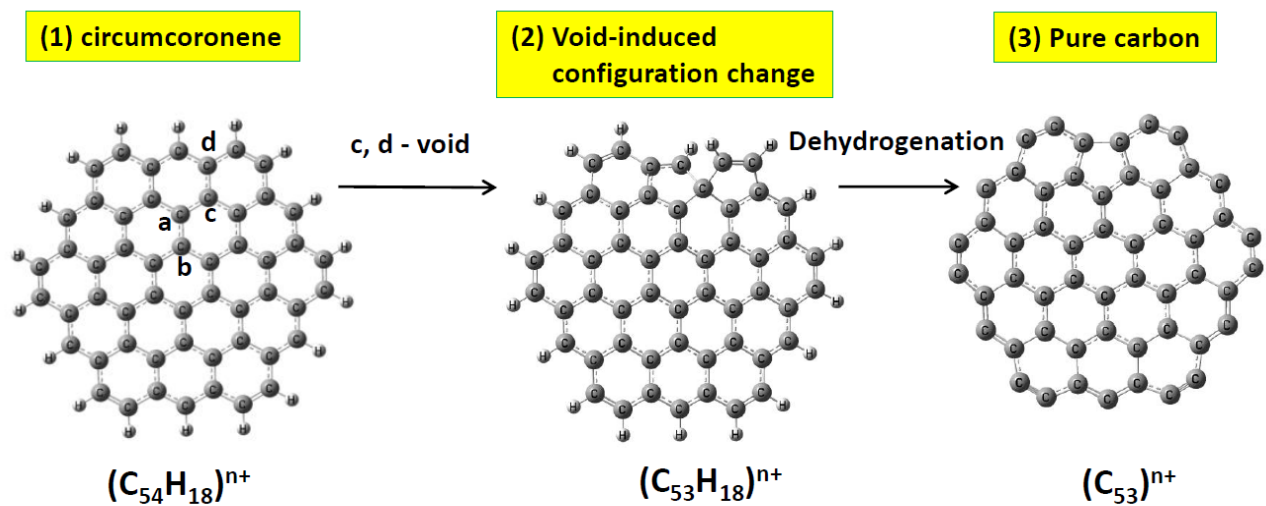

Figure 1, Calculation step of circumcoronene related molecules. (1) ionized cationic circumcoronene $(C_{54}H_{18})^{n+}$ (n=0, +1, +2. +3), (2) void at a carbon (position c or d) brings configuration change to $(C_{53}H_{18})^{n+}$ by Jahn-Teller effect, (3) Dehydrogenated pure carbon molecule $(C_{53})^{n+}$.

### 3, CALCULATION METHOD

In quantum chemistry calculation, we have to obtain total energy, optimized atom configuration, and infrared vibrational mode frequency and strength depend on a given initial atomic configuration, charge and spin state Sz. Density functional theory (DFT) with unrestricted B3LYP functional was applied utilizing Gaussian09 package (Frisch et al. 1984, 2009) employing an atomic orbital 6-31G basis set. The first step calculation is to obtain the self-consistent energy, optimized atomic configuration and spin density. Required convergence on the root mean square density matrix was less than $10^{-8}$ within 300 cycles. Based on such optimized results, harmonic vibrational frequency and strength was calculated. Vibration strength is obtained as molar absorption coefficient ε (km/mol.). Comparing DFT harmonic wavenumber $N_{DFT}$ (cm$^{-1}$) with experimental data, a single scaling factor 0.965 was used (Ota 2015b). Concerning a redshift for the anharmonic correction, in this paper we did not apply any correction to avoid over-estimation in a wide wavelength representation from 2 to 30 micrometer.

Corrected wave number N is obtained simply by N (cm$^{-1}$) = $N_{DFT}$ (cm$^{-1}$) x 0.965.

Wavelength λ is obtained by λ (micrometer) = 10000/N(cm$^{-1}$).

Reproduced IR spectrum was illustrated in a figure by a decomposed Gaussian profile with full width at half maximum FWHM=4cm$^{-1}$.

### 4, TYPE-A INFRARED SPECTRUM

Infrared spectrum of ionized circumcoronene was calculated as shown in Figure 2. Dashed vertical red lines are typically observed wavelength of 3.3, 6.2, 7.7, 8.6, 11.3, and 12.7 micrometer. Comparing with calculated blue curves, neutral circumcoronene $(C_{54}H_{18})$ shows good coincidence in wavelength, but no good in intensity ratio. Especially intensity at 11.1 micrometer is very large. Such a nature was similar with Type-A spectrum typically observed in astronomical object NGC1316 reported by B. Asabere et al. (Asabere 2016) and NGC4589 by J. D. Bregman et al. (Bregman 2006). In case of previously studied coronene related molecules, neutral coronene $(C_{24}H_{12})$ and pure carbon $(C_{23})^{2+}$ could reproduce such behavior.

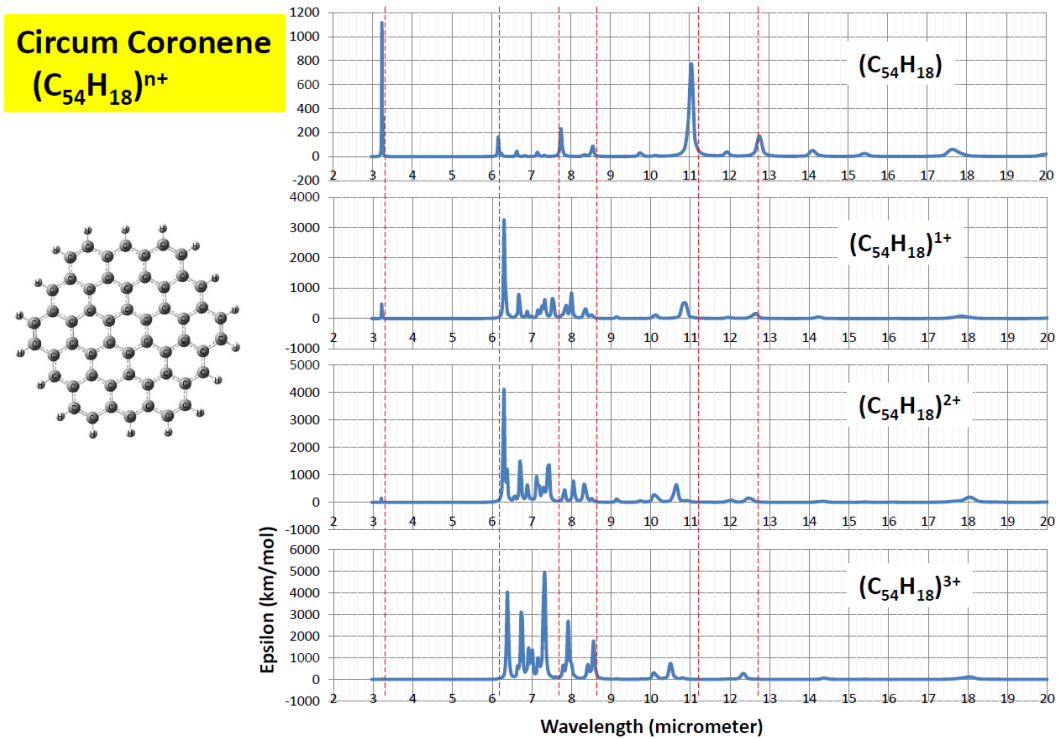

Figure 2, Calculated infrared spectrum of circumcoronene $(C_{54}H_{18})^{n+}$.

Figure 3, Type-A spectrum. Calculated spectrum of $(C_{54}H_{18})$ and $(C_{23})^{2+}$ are compared with observed one of NGC1316. Blue dashed lines coincident with both molecules, whereas green with only $(C_{54}H_{18})$, black with $(C_{23})^{2+}$, pink no coincidence with both molecules.

In Figure 3, we can compare calculated spectrum of ($C_{54}H_{18}$) and ($C_{23}$)$^{2+}$ with observed one of NGC1316. Blue dashed lines show good coincidence with above two molecules, especially very large peak strength at 11.3 micrometer band. Whereas, green dashed lines show coincidence with only ($C_{54}H_{18}$), also black dashed lines with only ($C_{23}$)$^{2+}$, and a pink dashed line at 9.6 micrometer show no coincidence with any. Reasonable idea is that observed Type-A spectrum may be a suitable mixture of those molecule candidates.

## 5, TYPE-B INFRARED SPECTRUM

Infrared spectrum of void induced molecules ($C_{53}H_{18}$)$^{n+}$ were calculated as shown in Figure 4. There are two hydrocarbon pentagons combined with 17 hexagons. Side view show Y-shape configuration. Among them, mono-cation one (n=1) revealed good coincidence with observed Type-B infrared spectrum. Typical astronomical objects are galaxy NGC6946 (Sakon 2007) and the red rectangle nebula (Mulas 2006). As illustrated in Figure 5, ubiquitously well observed 6.2, 7.7, 8.6, 11.3, and 12.7 micrometer bands could be reproduced fairly well by ($C_{53}H_{18}$)$^{1+}$. Detail is somewhat different that observed 11.2 micrometer band shifted to 10.9 micrometer in calculation. Whereas, observed 6.2 micrometer band was just reproduced by calculation. For comparison, previously calculated result of ($C_{23}H_{12}$)$^{2+}$ was illustrated on bottom (Ota 2017a).

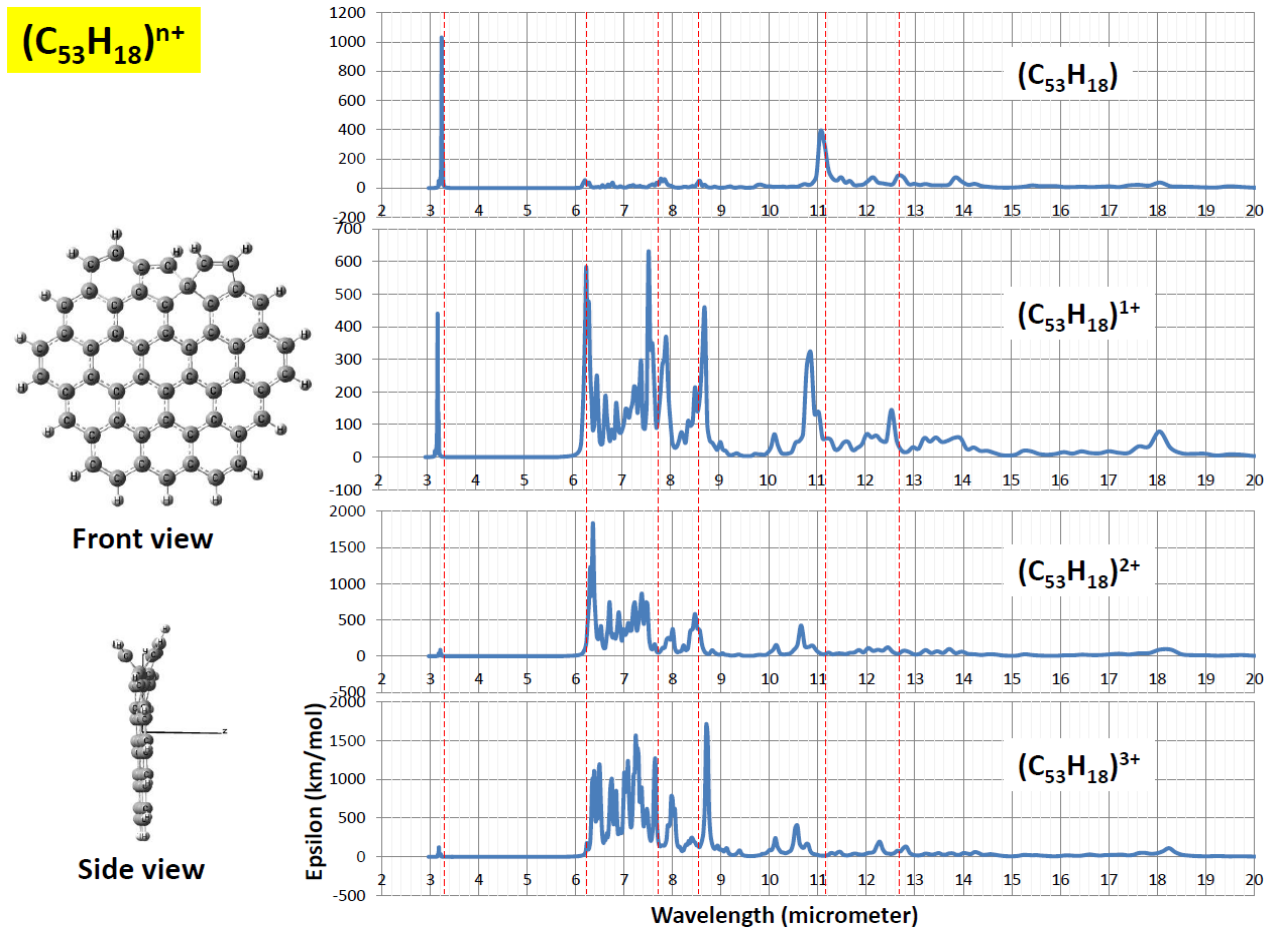

Figure 4, Calculated infrared spectrum of void induced circumcoronene ($C_{53}H_{18}$)$^{n+}$ (n=0, +1, +2, and +3).

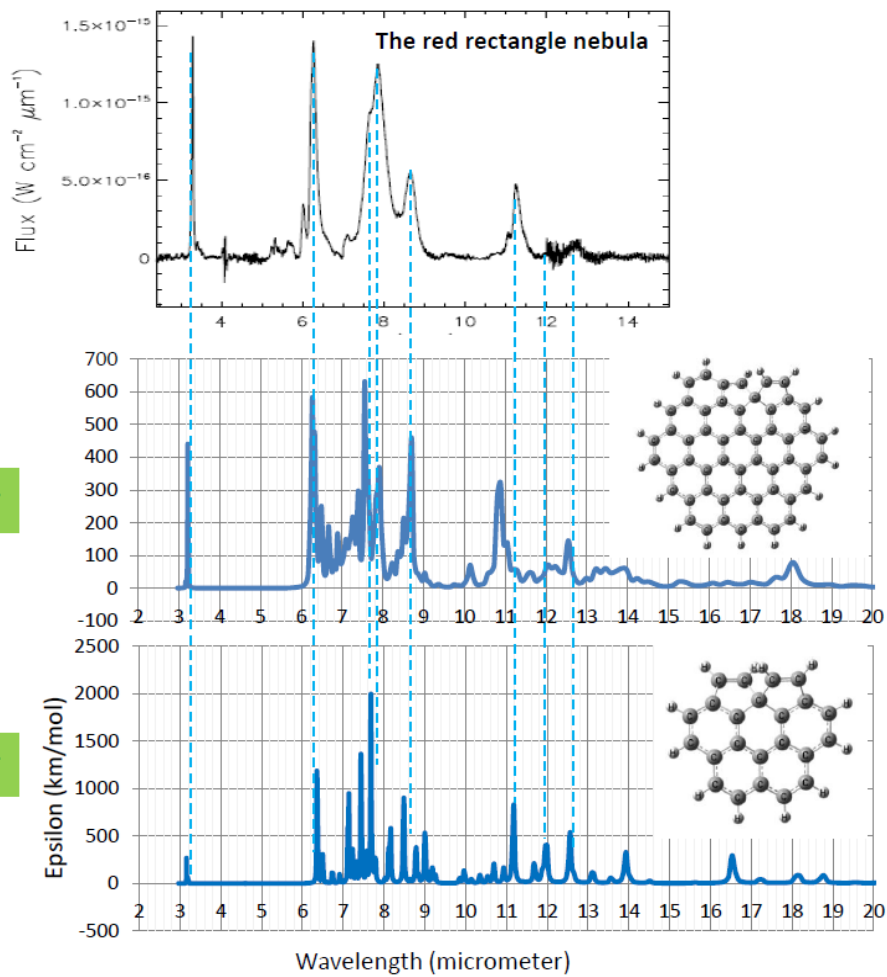

Figure 5, Type-B spectrum. On top, typical observed spectrum of the red rectangle. Middle panel is a calculated spectrum of void induced circumcoronene $(C_{53}H_{18})^{1+}$. Bottom is previously calculated result of void induced coronene $(C_{23}H_{12})^{2+}$.

6, TYPE-C INFRARED SPECTRUM

Most common ubiquitously observed spectrum was categorized as Type-C. Observed examples are NGC7023 (Boersma 2013), NGC2023 (Peeters 2017) and M17SW (Yamagishi 2016). Figure 6 show a typical example of NGC2023, where major bands are 6.2, 7.7, 8.6, 11.3, and 12.7 micrometer. Feature is very high peak at 11.3 micrometer than a height of 6.2 micrometer band. It should be reminded that in Type-A, 11.3 micrometer band is remarkably strong. One idea of Type-C is a mixture of Type-B and Type-A. Figure 6 also show calculated spectrum of $(C_{53}H_{18})^{1+}$ as Type-B and neutral circumcoronene $(C_{54}H_{18})$ as Type-A. If interstellar dust cloud would include those two carriers in some content, we can easily estimate an increase of 11.3 micrometer peak. In case of coronene related molecule, there are two candidates of Type-A, which were neutral coronene $(C_{24}H_{12})$ and pure carbon $(C_{23})^{2+}$. Unfortunately, in case of circumcoronene, we could not find a pure carbon candidate showing Type-A characteristic.

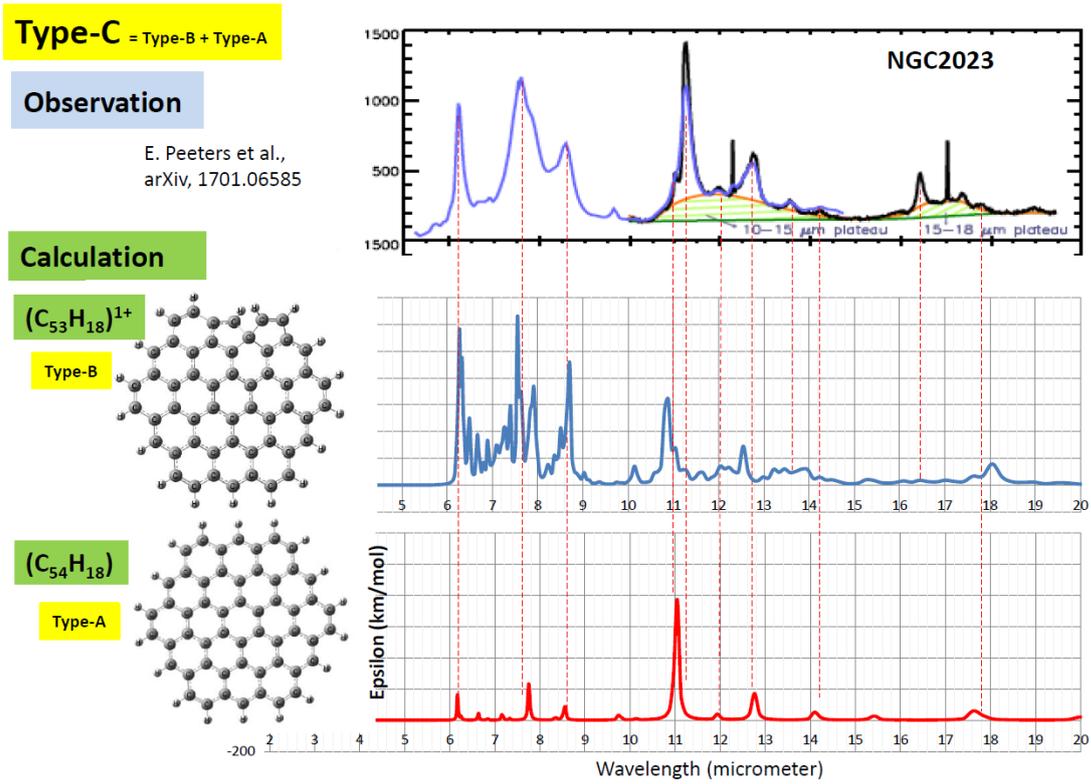

Figure 6, Type-C is common and ubiquitously observed in the universe. Observed examples is NGC2023 as illustrated on top. Feature is very high peak at 11.3 micrometer. Type-C could be explained by a mixture of Type-B and Type-A. In a series of circumcoronene, Type-B is $(C_{53}H_{18})^{1+}$, and Type-A $(C_{54}H_{18})$.

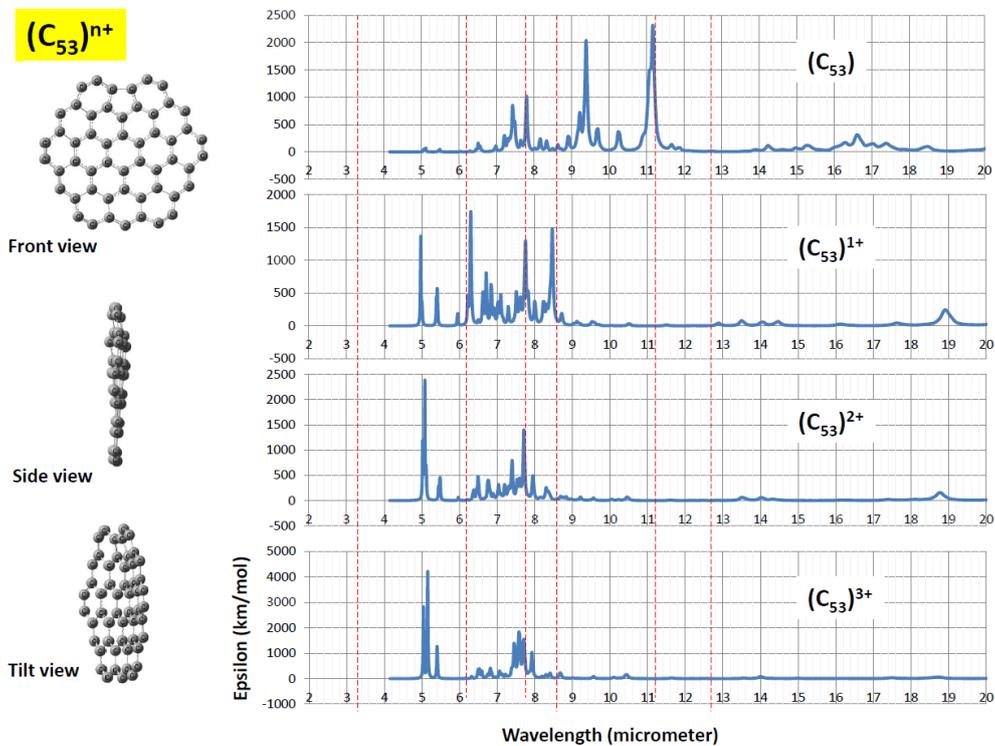

Figure 7, Calculated infrared spectrum of dehydrogenated pure carbon molecule $(C_{53})^{n+}$.

## 7, TYPE-E INFRARED SPECTRUM

Infrared spectrum of dehydrogenated molecule $(C_{53})^{n+}$ was shown in Figure 7. Due to strong tension of radical carbon network, molecule became a shallow cup like configuration looking at side view and tilt view. Among them, mono-cation $(C_{53})^{1+}$ predicted fairly well coincidence with well observed 6.2, 7.7, and 8.6 micrometer wavelength. However, there are no peaks at 11.3 micrometer. We like to name such spectrum to be Type-E. In Figure 8, paying attention on 11.3 micrometer band intensity (green broken line), we compared three types. Type-E has no peak, Type-B show medium height, whereas Type-A very strong peak. Actual interstellar dust cloud may include a mixture of those three types. Of course, carrier candidates are not only limited by circumcoronene, but also include coronene and other molecules. There are some examples showing smaller peak at 11.3 micrometer (Acke 2010). As illustrated in Figure 9, HD37411 predicted a capability of mixture of Type-B and Type-E.

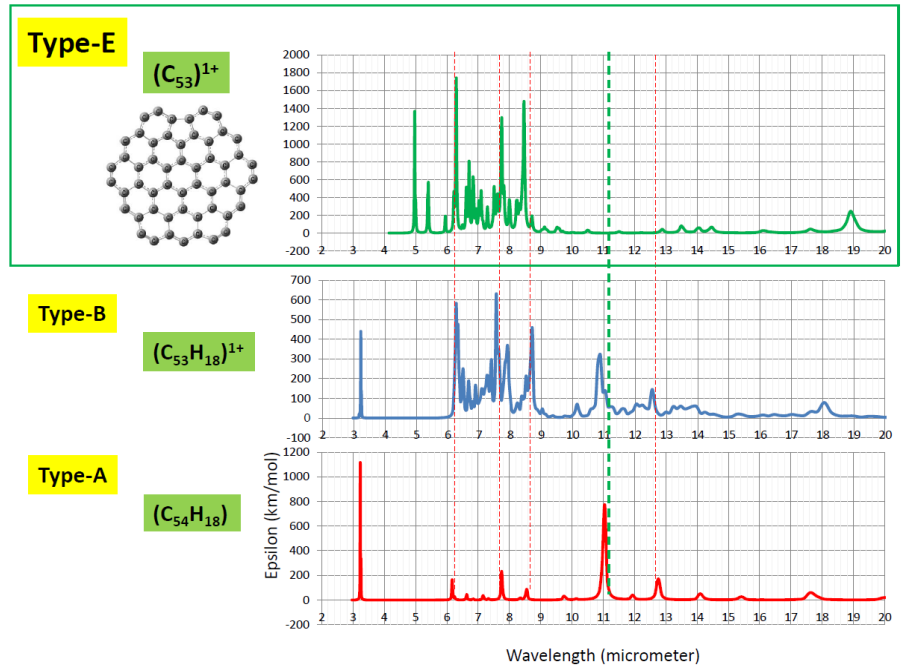

Figure 8, Comparison of spectrum of Type-E, Type-B, and Type-A

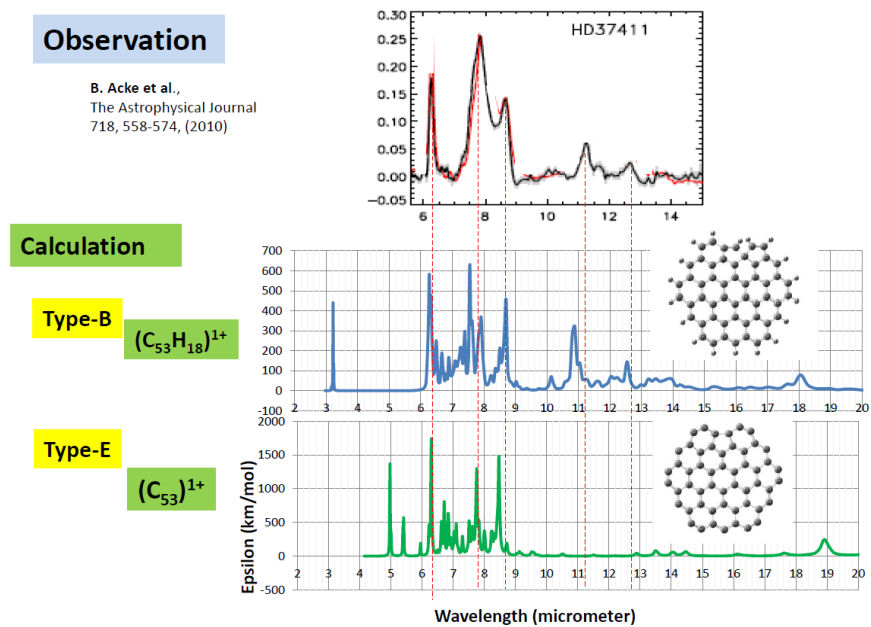

Figure 9, Spectrum of HD37411, which could be explained by a mixture of Type-B and Type-E.

## 8, BAND INTENSITY RATIO

In order to identify carrier molecules and to explain variation of spectrum, analysis of band intensity ratio is very important. Based on quantum chemical vibrational mode analysis, calculated intensity ratio was obtained as follows.
(1) Band intensity is sum of fundamental modes include main peak and peripheral sub peaks within plus/minus 0.1 micrometer. For example, 11.3 micrometer band is noted as P(11.3).
(2) Intensity band ratio between major bands was obtained as like P(7.7)/P(11.3)
(3) In order to compare with observed ratio, for example PAH(7.7)/PAH(11.3), scaling parameter "k" was introduced as like PAH(7.7)/PAH(11.3)=k x P(7.7)/P(11.3).
(4) In case of a mixture of Type-B and Type-A, mixture degree parameter "r" was introduced. An example is that,
   PAH(7.7)/PAH(11.3) = [ PAH(7.7)/PAH(11.3) of Type-B] + r x [PAH(7.7)/PAH(11.3) of Type-A]
  Also in case of a mixture of Type-B and Type-E,
   PAH(7.7)/PAH(11.3) = [ PAH(7.7)/PAH(11.3) of Type-B] + s x [PAH(7.7)/PAH(11.3) of Type-E]
(5) Other intensity ratio were also calculated as like,
   PAH(8.6)/PAH(11.3), PAH(6.2)/PAH(11.3), and PAH(12.7)/PAH(11.3)
Infrared intensity ratio based on calculated spectrum peaks of circumcoronene series were noted in Table 1.

Table 1, IR intensity ratio from calculated peaks

|  | $(C_{53}H_{18})^{1+}$ + $r[(C_{54}H_{18})]$ (Type-B) + r(Type-A) | | | $(C_{53}H_{18})^{1+}$ (Type-B) | $(C_{53}H_{18})^{1+}$ + $s[(C_{53})^{1+}]$ (Type-B) + s(Type-E) | | |  |
|---|---|---|---|---|---|---|---|---|
|  | r=1.0 | r=0.6 | r=0.2 |  | s=0.2 | s=0.3 | s=0.5 |  |
| PAH7.7/PAH11.2 | 1.23 | 1.60 | 2.44 | 3.43 (k=1.0) | 4.59 | 5.17 | 6.33 | X-axis |
| PAH8.6/PAH11.2 | 0.18 | 0.23 | 0.35 | 0.50 (k=0.43) | 0.93 | 1.14 | 1.82 | y1-axis |
| PAH6.2/PAH11.2 | 0.47 | 0.62 | 0.94 | 1.34 (k=0.49) | 1.94 | 2.25 | 2.85 | y2-axis |
| PAH12.7/PAH11.2 | 0.31 | 0.35 | 0.42 | 0.50 (k=1.20) | 0.52 | 0.52 | 0.53 | y3-axis |

### 8.1, Intensity ratio of PAH(8.6)/PAH(11.3)
   Figure 10 show intensity ratio of [PAH(8.6)/ PAH(11.3)] versus [PAH(7.7)/PAH(11.3)]. Observed spatial variation of M17SW (Yamagishi 2016) was illustrated by many small black dots. Also, galaxy scale observation results were overlapped by triangles and circles (Sakon 2007). Calculated result of Type-B $(C_{53}H_{18})^{1+}$ was marked by blue oval, which positioned almost mid of observed marks. Calculated line of a mixture of Type-B and Type-E is shown by a green broken line, which just trace M83 arm region (blue triangle) and NGC6946 arm region (blue circle). Also, calculated line of a mixture of Type-B and Type-A is drawn by a red broken line, which could reproduce M83 interarm (red triangle) and NGC6946 interarm region (red circle). Those calculated lines also traces upward top of M17SW observed dots (black small dots).

### 8.2, Intensity ratio of PAH(6.2)/PAH(11.3)
   Figure 11 illustrates intensity ratio of [PAH(6.2)/ PAH(11.3)] versus [PAH(7.7)/PAH(11.3)]. Observed marks are same with Figure 9. Calculated result of Type-B $(C_{53}H_{18})^{1+}$ was marked by blue oval, which positioned almost left end of M17SW spatial variation dots. Calculated line of a mixture of Type-B and Type-E is shown by a green broken line, which just trace NGC6946 arm region and NGC6946 LDGsource47. Also, calculated line of a mixture of Type-B and Type-A is drawn by a red broken line, which could reproduce M83 interarm (red triangle) and NGC6946 interarm region (red circle).

### 8.3, Intensity ratio of PAH(12.7)/PAH(11.3)

Figure 12 show intensity ratio of [PAH(12.7)/ PAH(11.3)] versus [PAH(7.7)/PAH(11.3)]. Black small dots are observed spatial variation of M17SW. Calculated result of Type-B $(C_{53}H_{18})^{1+}$ was marked by blue oval, which positioned almost left end of M17SW. Calculated line of a mixture of Type-B and Type-E is shown by a green broken line, which traces down end of spatial variation of M17SW. Also, calculated line of a mixture of Type-B and Type-A is drawn by a red broken line.

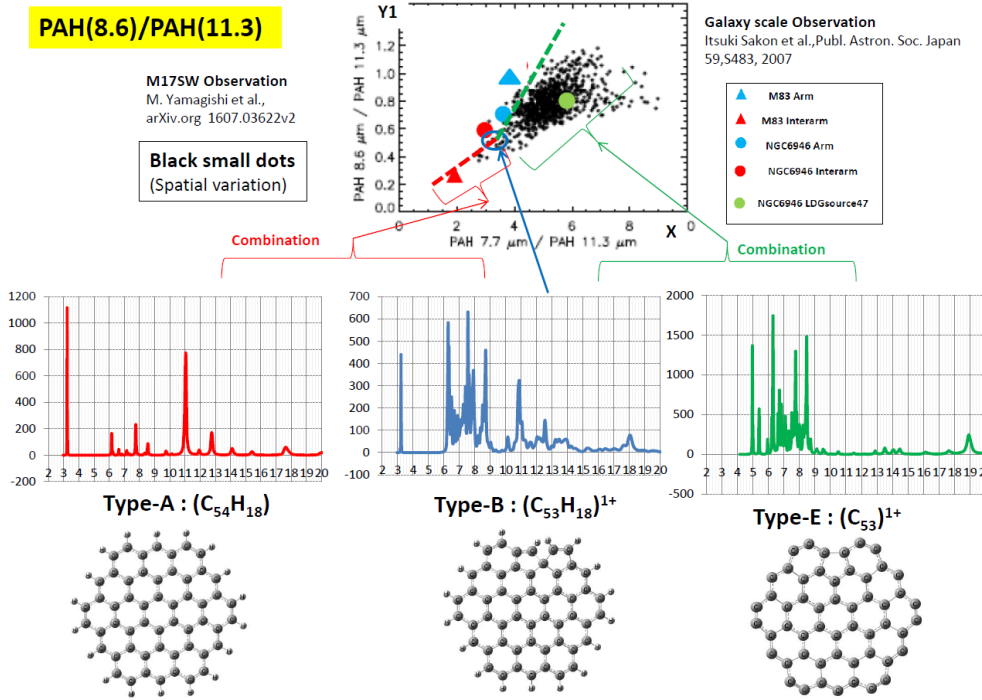

Figure 10, Intensity ratio of [PAH(8.6)/PAH(11.3)] versus [PAH(7.7)/PAH(11.3)]. Observed intensity could be explained by a mixture of Type-B, and Type-A, and Type-E.

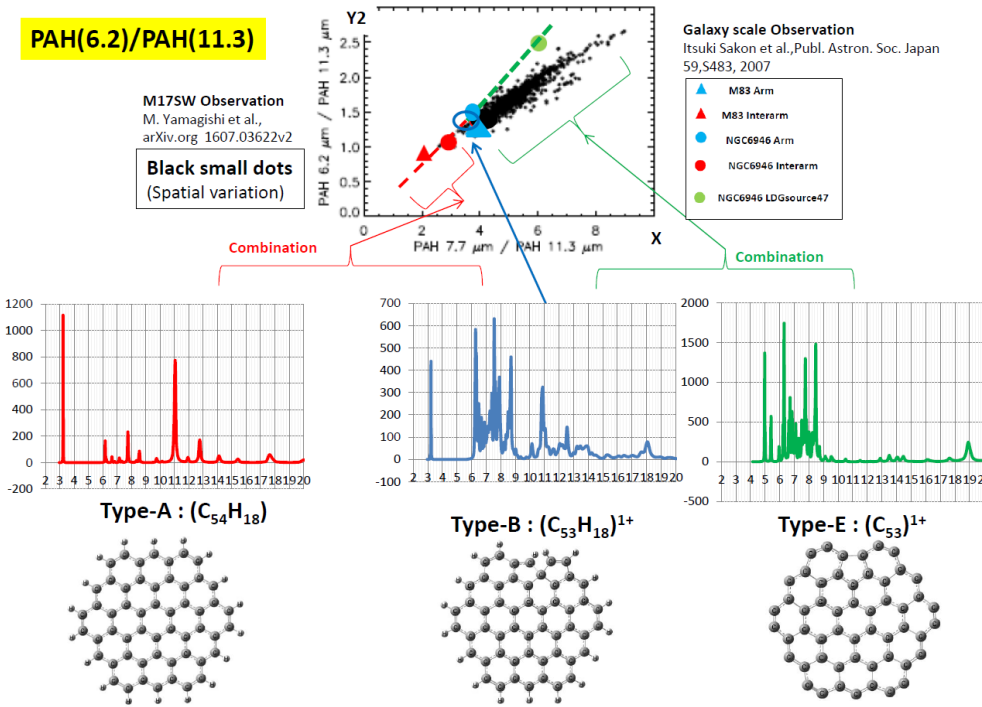

Figure 11, Intensity ratio of [PAH(6.2)/PAH(11.3)] versus [PAH(7.7)/PAH(11.3)]. Observed intensity could be explained by a mixture of Type-B, and Type-A, and Type-E.

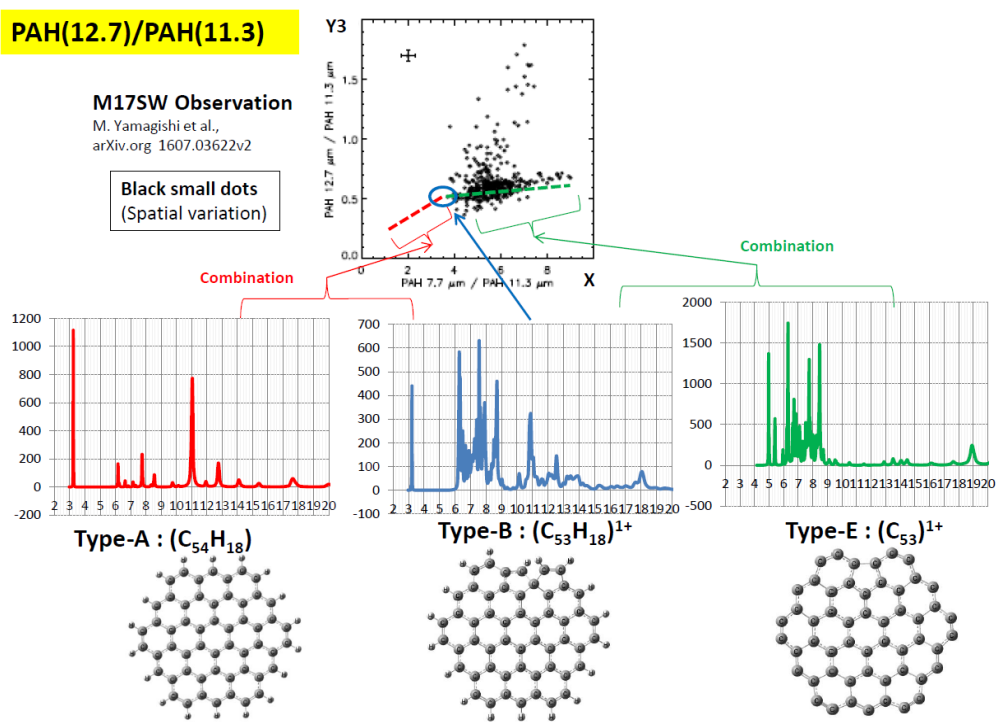

Figure 12, Intensity ratio of [PAH(12.7)/PAH(11.3)] versus [PAH(7.7)/PAH(11.3)]. Observed intensity could be explained by a mixture of Type-B and Type-E.

7, CONCLUSION

Identification of carrier molecules of astronomical interstellar infrared spectrum (IR) is important to understand chemical evolution step of polycyclic aromatic hydrocarbon (PAH). In our previous study, coronene ($C_{24}H_{12}$) related PAH could reproduce such IR by the first principles quantum chemical calculation. In this paper, PAH candidates were enhanced to circumcoronene ($C_{54}H_{18}$) related molecules.

(1)Type-B spectrum

Well known PAH oriented wavelength of 6.2, 7.7, 8.6, 11.3, and 12.7 micrometer were reproduced well by a void induced mono-cation $(C_{53}H_{18})^{1+}$ having hydrocarbon two pentagons combined with 17 hexagons. Typical astronomical object are the red rectangle nebula and NGC6946.

(2) Type-A spectrum

Neutral circumcoronene ($C_{54}H_{18}$) show very strong IR peak at 11.1 micrometer, which could reproduces main part of Type-A spectrum observed in NGC1316 and NGC4589.

(3) Type-C spectrum

Ubiquitously well observed spectrum was noted as Type-C, which could be explained by a suitable combination of Type-B and Type-A. In case of circumcoronene related molecules, Type-C would be suitable sum of ($C_{54}H_{18}$) and $(C_{53}H_{18})^{1+}$. Typical astronomical objects are NGC7023, NGC2023 and so many.

(4) Type-E spectrum

Dehydrogenated pure carbon mono-cation molecule $(C_{53})^{1+}$ show IR peaks at 6.3, 7.8, 6.5 micrometer, but no peak at 11.3 micrometer, which newly defined as Type-E. Such a strange characteristic can contribute to explain IR intensity ratio.

(5) Intensity ratio

Observed intensity ratio between peaks of 6.2, 7.7, 8.6, 11.3, and 12.7 micrometer each other were compared with above calculated IR intensities. Major intensity ratio was well reproduced by Type-B molecules. Variation of observed intensity ratio could be explained by a mixture degree between Type-B and Type-A, and also a mixture degree between Type-B and Type-E.

ACKNOWLEDGEMENT

I would like to say great thanks to Prof. Takashi Onaka and Prof. Itsuki Sakon, University of Tokyo, for very kind discussion on infrared astronomy, especially to trigger me an importance of enhancing molecular size and intensity ratio.

Author profile
  Norio Ota PhD.
    Senior Professor, University of Tsukuba, Japan
    Material Science,
    Optical data storage

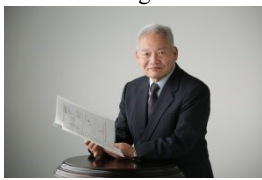